\documentstyle[12pt]{article}
\def\hybrid{\topmargin -20pt	\oddsidemargin 0pt
	\headheight 0pt	\headsep 0pt
	\textwidth 6.25in	
	\textheight 9.5in	
	\marginparwidth .875in
	\parskip 5pt plus 1pt	\jot = 1.5ex}

\hybrid

\def\baselinestretch{1.2}

\catcode`\@=11

\def\marginnote#1{}
\newcount\hour
\newcount\minute
\newtoks\amorpm
\hour=\time\divide\hour by60
\minute=\time{\multiply\hour by60 \global\advance\minute by-\hour}
\edef\standardtime{{\ifnum\hour<12 \global\amorpm={am}%
	\else\global\amorpm={pm}\advance\hour by-12 \fi
	\ifnum\hour=0 \hour=12 \fi
	\number\hour:\ifnum\minute<10 0\fi\number\minute\the\amorpm}}
\edef\militarytime{\number\hour:\ifnum\minute<10 0\fi\number\minute}
\def\draftlabel#1{{\@bsphack\if@filesw {\let\thepage\relax
   \xdef\@gtempa{\write\@auxout{\string
      \newlabel{#1}{{\@currentlabel}{\thepage}}}}}\@gtempa
   \if@nobreak \ifvmode\nobreak\fi\fi\fi\@esphack}
	\gdef\@eqnlabel{#1}}
\def\@eqnlabel{}
\def\@vacuum{}
\def\draftmarginnote#1{\marginpar{\raggedright\scriptsize\tt#1}}

\def\draft{\oddsidemargin -.5truein
	\def\@oddfoot{\sl preliminary draft \hfil
	\rm\thepage\hfil\sl\today\quad\militarytime}
	\let\@evenfoot\@oddfoot	\overfullrule 3pt
	\let\label=\draftlabel
	\let\marginnote=\draftmarginnote
   \def\@eqnnum{(\theequation)\rlap{\kern\marginparsep\tt\@eqnlabel}%
\global\let\@eqnlabel\@vacuum}  }

\def\preprint{\twocolumn\sloppy\flushbottom\parindent 2em
	\leftmargini 2em\leftmarginv .5em\leftmarginvi .5em
	\oddsidemargin -.5in	\evensidemargin -.5in
	\columnsep .4in	\footheight 0pt
	\textwidth 10.in	\topmargin  -.4in
	\headheight 12pt \topskip .4in
	\textheight 6.9in \footskip 0pt
	\def\@oddhead{\thepage\hfil\addtocounter{page}{1}\thepage}
	\let\@evenhead\@oddhead	\def\@oddfoot{}	\def\@evenfoot{} }

\def\numberbysection{\@addtoreset{equation}{section}
	\def\theequation{\thesection.\arabic{equation}}}

\def\underline#1{\relax\ifmmode\@@underline#1\else
	$\@@underline{\hbox{#1}}$\relax\fi}

\def\titlepage{\@restonecolfalse\if@twocolumn\@restonecoltrue\onecolum
n
     \else \newpage \fi \thispagestyle{empty}\c@page\z@
	\def\thefootnote{\fnsymbol{footnote}} }

\def\endtitlepage{\if@restonecol\twocolumn \else \newpage \fi
	\def\thefootnote{\arabic{footnote}}
	\setcounter{footnote}{0}}  

\catcode`@=12
\relax

\def\figcap{\section*{Figure Captions\markboth
	{FIGURECAPTIONS}{FIGURECAPTIONS}}\list
	{Figure \arabic{enumi}:\hfill}{\settowidth\labelwidth{Figure
999:}
	\leftmargin\labelwidth
	\advance\leftmargin\labelsep\usecounter{enumi}}}
 \relax
\def\tablecap{\section*{Table Captions\markboth
	{TABLECAPTIONS}{TABLECAPTIONS}}\list
	{Table \arabic{enumi}:\hfill}{\settowidth\labelwidth{Table
999:}
	\leftmargin\labelwidth
	\advance\leftmargin\labelsep\usecounter{enumi}}}
 \relax
\def\reflist{\section*{References\markboth
	{REFLIST}{REFLIST}}\list
	{[\arabic{enumi}]\hfill}{\settowidth\labelwidth{[999]}
	\leftmargin\labelwidth
	\advance\leftmargin\labelsep\usecounter{enumi}}}
 \relax
\makeatletter
\newcounter{pubctr}
\def\publist{\@ifnextchar[{\@publist}{\@@publist}}
\def\@publist[#1]{\list
	{[\arabic{pubctr}]\hfill}{\settowidth\labelwidth{[999]}
	\leftmargin\labelwidth
	\advance\leftmargin\labelsep
	\@nmbrlisttrue\def\@listctr{pubctr}
	\setcounter{pubctr}{#1}\addtocounter{pubctr}{-1}}}
\def\@@publist{\list
	{[\arabic{pubctr}]\hfill}{\settowidth\labelwidth{[999]}
	\leftmargin\labelwidth
	\advance\leftmargin\labelsep
	\@nmbrlisttrue\def\@listctr{pubctr}}}
 \relax
\makeatother
\newskip\humongous \humongous=0pt plus 1000pt minus 1000pt

\newif\ifdtup

\relax

\def\s{\sigma}
\def\thefootnote{\fnsymbol{footnote}}
\def\be{\begin{equation}}
\def\ee{\end{equation}}
\def\ba{\begin{eqnarray}}
\def\ea{\end{eqnarray}}
\def\d{\partial}
\def\db{\bar{\partial}}
\def\G{\Gamma}
\def\S{\Sigma}
\def\Jb{\bar{J}}
\def\a{\alpha}
\def\th{\theta}
\def\tha{\theta_1}
\def\thb{\theta_2}
\def\tht{\tilde{\theta}}
\def\Jb{\bar{J}}

\def\p{\phi}
\def\e{\epsilon}
\begin{document}
\renewcommand{\theequation}{\thesection.\arabic{equation}}
\newcommand{\beq}{\begin{equation}}
\newcommand{\eeq}[1]{\label{#1}\end{equation}}
\newcommand{\ber}{\begin{eqnarray}}
\newcommand{\eer}[1]{\label{#1}\end{eqnarray}}
\begin{titlepage}
\begin{center}

\hfill CERN-TH.6816/93\\
\hfill RI-149-93\\
\hfill hep-th/9303016\\

\vskip .5in

{\large \bf Axial-Vector Duality as a Gauge Symmetry and Topology
Change
in String Theory}
\vskip .5in

{\bf Amit Giveon} \footnotemark \\

\footnotetext{e-mail address: GIVEON@HUJIVMS.bitnet}

\vskip .1in

{\em Racah Institute of Physics\\ The Hebrew University\\ Jerusalem,
91904,
ISRAEL}

\vskip .15in

       and

\vskip .15in

{\bf Elias Kiritsis} \footnote{e-mail address:
KIRITSIS@NXTH08.CERN.CH}\\
\vskip
 .1in

{\em Theory Division, CERN, CH-1211\\
Geneva 23, SWITZERLAND}\\

\vskip .1in

\end{center}

\vskip .4in

\begin{center} {\bf ABSTRACT } \end{center}
\begin{quotation}\noindent
Lines generated by marginal deformations of WZW models are
considered. The
Weyl symmetry at the WZW point implies the existence of a duality
symmetry
on such lines. The duality  is interpreted as a broken gauge symmetry
in
string theory. It is shown that at the two end points   the axial and
vector cosets are obtained. This shows that the axial and vector
cosets are equivalent CFTs both in the compact and the non-compact
cases.
Moreover, it is shown that there are $\s$-model deformations that
interpolate smoothly
between manifolds with different topologies.
\end{quotation}
\vskip1.0cm
CERN-TH.6816/93 \\
February 1993\\
Revised: August 1993\\
\end{titlepage}
\vfill
\eject
\def\baselinestretch{1.2}
\baselineskip 16 pt
\noindent
\section{Introduction}
\setcounter{equation}{0}

The moduli space of flat d-dimensional toroidal string
compactifications
is generated by a group $O(d,d,R)$ \cite{nsw}.
This generalizes to curved string backgrounds with $d$ toroidal
isometries
(chiral and anti-chiral) where the moduli space again can be
generated by the
action
of a group
isomorphic to $O(d,d,R)$ \cite{moduli,GR}. The global structure of
the
moduli space is rather striking; points corresponding to different
backgrounds, are related by the generalized duality group of discrete
symmetries, isomorphic to $O(d,d,Z)$ \cite{GRV,GR}.

In this paper, we will argue that  the discrete symmetries of the
moduli of
target space should be interpreted as global residual symmetries of
an
underlying broken gauge group of string theory. In particular, we
will
show that the axial/vector duality of abelian WZW cosets \cite{K} is
a
gauge symmetry, and therefore, it is an exact symmetry in string
theory,
both for compact as well as non-compact groups.

Moreover, we will describe deformations that generate a family of
conformal
backgrounds that interpolate between manifolds with
different topologies. This provides a simple example of classical
topology
change in string theory, similar to the phenomenon described recently
for
Calabi-Yau compactifications \cite{Getal,Witten}. The  topology
change that
will be described here occurs for cosmological string backgrounds,
and is
therefore important.

The view of discrete symmetries in target space as gauge symmetries
was
already discussed in the flat case. In ref. \cite{DHS} it was argued
that
from the point of view of an effective space-time theory, the
discrete
symmetry relating a compact dimension of radius $r$ to that of radius
$1/r$
is indeed a part of a continuous $SU(2)$ rotation. The argument goes
as
follows. When the compactified circle is at radius $r=1$, there is an
extended affine symmetry $SU(2)_L\times SU(2)_R$. This theory
contains
three chiral currents $J^a$, and three anti-chiral currents $\Jb^a$
in the
adjoint representation of $SU(2)_L$ and $SU(2)_R$, respectively.  The
space
of truly marginal independent directions is embedded in the
9-dimensional
space generated by $J^a\Jb^b$. The set of critical points that can be
reached from the $SU(2)$ point by conformal deformations (of the type
$(\sum_{a=1}^{a=3}\a_a J^a)(\sum_{b=1}^{b=3}\beta_b \Jb^b)$) span a
5-dimensional surface in the 9-dimensional euclidean space.  However,
because different truly marginal perturbations are equivalent under
continuous transformations in the group $SU(2)_L\times SU(2)_R$, the
dimension of the physical moduli space is 1.  In particular, the
duality
transformation relating $r$ to $1/r$ corresponds to the Weyl
transformation
in $SU(2)_L$ that takes $J^3$ to $-J^3$. Infinitesimally near the
$SU(2)$
point, this corresponds to the identification of the theory given by
the
deformation $\a J^3\Jb^3$, with the theory given by the deformation
$-\a
J^3\Jb^3$.

The generalization to $d$-dimensional toroidal backgrounds was given
ref.
\cite{GMR2}. It was shown that any element of the generalized duality
group
$O(d,d,Z)$, is a product of discrete symmetries corresponding to
continuous
rotations in groups (Weyl transformations) around points with an
extended
affine symmetry. Moreover, it was suggested \cite{GMR2} that in
string
theory the generalized dualities are residual discrete symmetries of
a
broken infinite dimensional gauge group. This was realized for the
heterotic string compactified on a torus in ref. \cite{GP}, where a
completely duality invariant effective action of the $N=4$ heterotic
string
was constructed. It was then shown that the infinite dimensional
gauge
algebra is broken at any classical background to a finite dimensional
group, and that the generalized duality group $O(6,22,Z)$ elements
are
residual discrete symmetries of the broken gauge algebra. It is
therefore
expected that target space dualities are  residual  discrete
symmetries of
the underlying gauge algebra of string  theory
\cite{GMR2,GP,KZ,G,IL}.

The identification of target space dualities with continuous
rotations
around points with an extended symmetry is rather important. In
particular,
it proves that such dualities are exact symmetries (
nonperturbatively in $\alpha$'): theories associated to different
flat backgrounds that
are related by duality correspond to the same CFT (to all orders and
interactions).

Generalized target space dualities are not limited only to flat
backgrounds; in ref. \cite{GR} it was shown that the elements of
$O(d,d,Z)$
are  discrete symmetries of the space of curved string backgrounds
that are
independent of $d$ coordinates. It was moreover shown in \cite{K2}
that (as
in the flat case \cite{GMR2}), in the compact WZW case, these are
exact
symmetries of CFT and string theory  by relating them  to invariance
under
the affine Weyl group. It was also argued that they should persist in
the
non-compact case.

Are these symmetries residual discrete symmetries of an underlying
gauge
algebra in string theory? In this work we present the answer
concerning
particular elements of the $O(d,d,Z)$ group. Namely, we will show
that any
element relating an axially gauged $U(1)$ of a WZW model to the
vector
gauging is a residual discrete symmetry of the broken gauge algebra
in the
sense discussed above. In particular, this proves that the
axial/vector
coset duality is an exact symmetry in string theory. This is true for
compact groups, as well as for non-compact groups, and therefore, has
important implications for black-hole duality \cite{GDVV} and the
study of
singularities  in string theory. The extension to the full $O(d,d,Z)$
might
be done along the lines of \cite{GMR2}, and will appear elsewhere.

The paper is organized as follows:  In section 2 we start with the
simplest
non-trivial case,  the $SU(2)$ or $SL(2,R)$ model and its marginal
deformation. In section 3 we present the partition function, and
discuss
the geometrical interpretation of the modulus parameter. In section 4
we
study the target space geometry and topology change along the line of
deformations, and we describe a smooth topology change in the
extended
moduli space of the $SU(2)$ WZW model. In section 5 we deal with the
general case. In section 6 we discuss the duality on the line of
marginal
deformations and its relation to a broken gauge symmetry
transformation.
Finally, section 7 contains our conclusions and further comments.
In the appendix we describe why the $\sigma$-model action we give for
the
deformed WZW model is exact to all orders in $\a'$.

\section{$J\Jb$ deformation of $SU(2)$ or $SL(2)$ WZW model
and duality as a broken gauge symmetry}
\setcounter{equation}{0}

In this section we will consider duality as a broken gauge symmetry
for
the simplest nontrivial case, namely, duality acting on the
deformation
line of
$SU(2)$ or $SL(2)$ WZW models.\footnote{
We define the CFT corresponding to the $SL(2)$ WZW model as the
analytic
continuation of the (Euclidean) 3-d hyperboloid ($H^{+}_{3}$)
$\s$-model.
This definition provides with a stable path integral prescription
for the $SL(2)$ theory as shown in \cite{Ga}.
All subsequent remarks concerning $SL(2)$ will assume this
definition.}

If we parametrize the $SU(2)$ group element as
\be
g=e^{i\theta_{1}\s_{3}}e^{ix\s_{2}}e^{i\theta_{2}\s_{3}}
\ee
then the  action for $SU(2)_k$ WZW model is given
by
\ba
S[x,\tha,\thb]&=&S_1+S_a+S[x], \nonumber\\
S_1&=&\frac{k}{2\pi}\int d^2 z (\d\tha \db\tha + \d\thb \db\thb
+2\S (x) \d\thb \db\tha), \nonumber\\
S_a&=&\frac{k}{2\pi}\int d^2 z(\d\thb\db\tha -
\d\tha\db\thb),\nonumber\\
S[x]&=&\frac{k}{2\pi}\int d^2 z\; \d x \db x -
\frac{1}{8\pi}\int d^2 z\; \p_0 R^{(2)},
\label{su2}
\ea
where $\S(x)=\cos 2x$, and $\p_0$ is a constant dilaton.
The action for $SL(2)_k$ WZW model is given from (\ref{su2}) by
taking
$x\rightarrow ix$ and $k\rightarrow -k$.

The antisymmetric term $S_a$ in (\ref{su2}) is (locally) a total
derivative, and therefore may give only topological contributions,
depending on the periodicity of the coordinates $\th $.
To specify the periodicity, we define
\be
\th=\thb-\tha\; , \qquad \tht=\tha+\thb\; ,
\label{th}
\ee
such that
\be
\th\equiv \th+2\pi\; , \qquad  \tht\equiv\tht+2\pi\; .
\label{th1}
\ee
In these coordinates the action becomes
\be
S[x,\th,\tht]=\frac{1}{2\pi}\int d^2 z
(\d\th, \d\tht, \d x)\left( \begin{array}{clcr}
                                          E & 0\\
                                          0 & k
                                         \end{array}\right)
\left(\begin{array}{clcr} \db \th\\ \db\tht \\ \db x\end{array}
\right)
-\frac{1}{8\pi}\int d^2 z \p_0 R^{(2)},
\label{Ssu2}
\ee
where $E$ is the $2\times 2$ matrix
\be
E=\frac{k}{2}\left(\begin{array}{clcr}
        1-\S & 1+\S\\
      -(1+\S) & 1+\S \end{array}\right),
\label{Esu2}
\ee

The action $S$ (\ref{Ssu2}) is manifestly invariant under the
$U(1)_L\times U(1)_R$ affine symmetry generated by the currents
\ba
J=\frac{k}{2}(-(1-\S)\d\th +(1+\S)\d\tht)\; , \nonumber\\
\Jb=\frac{k}{2}((1-\S)\db\th +(1+\S)\db\tht)\; .
\label{JJ}
\ea
In addition, there are two extra chiral currents, and
anti-chiral currents, completing the affine $SU(2)_L\times SU(2)_R$
or $SL(2)_L\times SL(2)_R$
symmetry of the the WZW model.

It is possible to deform the action $S$ to new conformal backgrounds
by
adding to it any marginal  deformation. We will focus on marginal
deformations that are obtained as a linear combination of chiral
currents
times a linear combination of anti-chiral currents.\footnote{ It is
easy to
see that for generic level these are the only potential marginal
deformations. Any other possible deformation has to be generated by
affine
primaries, and these do not generically have dimension (1,1). This
can
happen though at special levels. For example, in the case of
$SU(2)_{k}$
when $k=(m-1)(m+2)$, then the $j=m$ primary has dimension one. It is
not
known if this is exactly marginal, although it is for $m=2$. However
it is
suggestive that the central charge of the associated parafermion
system is
$1+[1-6/m(m+1)]$ so that it may be that the theory is a semidirect
product
of a $U(1)$ and a minimal model.} It is now important to note that
all the
deformations that are equivalent under the action of the symmetry
group
give rise to equivalent CFTs (although not necessarily to backgrounds
that
are related by coordinate transformations!). In the following we
deform the
WZW action with the $J\Jb$ marginal operator, as was done in ref.
\cite{HS}. This deformation gives rise to a one parameter family of
theories, parametrized by the radius of the Cartan torus.

Once deforming with $J\Jb$, the affine symmetry is broken to
$U(1)_L\times
U(1)_R$. The $U(1)$ chiral and anti-chiral currents at the deformed
theory
can be found and will be presented below.

The $J\Jb$ deformation is equivalent to a particular one parameter
family
of $O(2,2)$ rotations acting on the background matrix $E$ and the
dilaton
\cite{HS}. To show this point we begin by establishing our notation
following ref. \cite{GMR2}.

The group $O(d,d,R)$ can be represented as a $2d\times
2d$-dimensional
matrices $g$ preserving the bilinear form $j$
\be
g=\left(\begin{array}{clcr} a&b\\c&d\end{array}\right), \qquad
j=\left(\begin{array}{clcr} 0&I\\I&0\end{array}\right),
\label{gJ}
\ee
where $a,b,c,d$, and $I$ are $d\times d$-dimensional constant
matrices, and
\be
g^tjg=j.
\ee
We define the action of $g$ on $E$ by fractional linear
transformations:
\be
g(E)=E'=(aE+b)(cE+d)^{-1}.
\label{gE}
\ee

The group $O(d,d,R)$ is generated  \cite{GMR2} by $GL(d)$
transformations:
\be
\left(\begin{array}{clcr} a&b\\c&d\end{array}\right)=
\left(\begin{array}{cccc} A&0\\0&(A^t)^{-1}\end{array}\right)
\qquad {\rm s.t.} \;\;\; A\in GL(d),
\ee
constant $\Theta$ shifts
\be
\left(\begin{array}{clcr} a&b\\c&d\end{array}\right)=
\left(\begin{array}{clcr} I&\Theta\\0&I\end{array}\right)
\qquad {\rm s.t.} \;\;\; \Theta=-\Theta^t,
\ee
and factorized duality
\be
\left(\begin{array}{clcr} a&b\\c&d\end{array}\right)=
\left(\begin{array}{cccc} I-e_1&e_1\\e_1&I-e_1\end{array}\right)
\qquad {\rm s.t.} \;\;\; e_1={\rm diag}(1,0,...,0).
\label{factd}
\ee
The discrete group $O(d,d,Z)$ is defined to be the elements $g$ in
(\ref{gJ}) with integer entries.

Rotations with $O(2,2)$ elements in $GL(2)$ and $\Theta$ shifts
preserve
the current algebra (although in general they change the spectrum) .
Therefore, once choosing a $J\Jb$ deformation of the WZW point
(\ref{su2})
for specific $J$ and $\Jb$, the deformed theories are described in
terms of
a one parameter family of rotations. Next we describe this rotation.

The backgrounds corresponding to the $J\Jb$ deformation of the WZW
point,
with $J$ and $\Jb$ given in (\ref{JJ}), are constructed by an
$O(2,2)$
transformation (\ref{gE}) of the WZW background (\ref{Esu2}) with the
element \cite{HS}
\be
g_{\a}=
\left(\begin{array}{cccc}
I&\cos^2\a(k-\tan\a)\e\\0&I\end{array}\right)
\left(\begin{array}{cccc} A(\a)&0\\0&(A(\a)^t)^{-1}\end{array}\right)
\left(\begin{array}{cccc} C(\a)&S(\a)\\S(\a)&C(\a)\end{array}\right)
\left(\begin{array}{cccc} I&-k\e\\0&I\end{array}\right),
\label{ga}
\ee
where
\ba
I=\left(\begin{array}{cccc} 1&0\\0&1\end{array}\right),\qquad
\e=\left(\begin{array}{cccc} 0&1\\-1&0\end{array}\right),&{}&\qquad
A(\a)=\left(\begin{array}{cccc} \cos\a&0\\0&\cos\a(1+k\tan\a)
\end{array}\right)\nonumber\\
C(\a)=\cos\a\; I, &{}&\qquad S(\a)=\sin\a\;\e,
\label{e}
\ea
namely,
\be
\frac{1}{k}g_{\a}(E)\equiv E_{R(\a)}=\frac{1}{1+R^2\frac{1-\S}{1+\S}}
\left(\begin{array}{cccc}
\frac{1-\S}{1+\S}&1\\-1&R^2\end{array}\right),
\label{ER}
\ee
where
\be
R(\a)^2=(1+k\tan\a)^2.
\label{R}
\ee
The deformation line parametrized by $-\pi/2\leq\a\leq\pi/2$ is,
therefore,
a double cover of the line parametrized by the radius $0\leq
R\leq\infty$.
The original WZW point is given at $R=1$.

The dilaton field $\p$ also transforms under the $O(2,2)$ rotation.
At the WZW
point the dilaton is the constant $\p=\p_0$ appearing in (\ref{su2}).
Then at the point $R$, the value of the dilaton is
equal to that implied by $O(2,2,R)$ tranformations (see also appendix
A)
\be
\p=\p_{0}+{1\over 2}\log\left({{\rm det}G(1)\over {\rm
det}G(R)}\right).
\label{pR}
\ee
With this dilaton, $\sqrt{G(R)}e^{\phi(R)}$ is independent of R, and
 moreover, it
has the appropriate asymptotic behaviour as $R\to\infty$.

We have obtained the $\sigma$-model background (\ref{ER},\ref{pR})
using O(2,2)
transformations, which are correct to 1-loop order (but correctable
to all
orders \cite{K2}).
In appendix A we show that there is  a scheme in which this
background
solves the $\beta$-function equations to all orders.

Two special backgrounds occur at the boundaries of the $R$ modulus
space.
At $R=0$ ($\a=tan^{-1}(-\frac{1}{k}$)) the background matrix is
\be
E_{R=0}=
\left(\begin{array}{cccc} \frac{1-\S}{1+\S}&1\\-1&0\end{array}\right)
=\left(\begin{array}{cccc} \frac{1-\S}{1+\S}&0\\0&0\end{array}\right)
\;\;\; {\rm mod} \;\; \Theta - {\rm shift}.
\label{pipa}
\ee
This background
corresponds to the direct product of the vectorially gauged coset
$SU(2)/U(1)_v$ \cite{brc} or $SL(2)/U(1)_v$ and a free
boson at a compactification
radius $r=0$ (that is equivalent to a non-compact free boson via the
$r\rightarrow 1/r$ duality).
The constant antisymmetric tensor in (\ref{pipa}) can be safely
dropped
since one of the two coordinates is non-compact.

At $R=\infty$ ($\a=\pi/2$) the background matrix is
\be
E_{R=\infty}=
\left(\begin{array}{cccc} 0&0\\
0&\frac{1+\S}{1-\S}\end{array}\right).
\ee
This background corresponds to the direct product of the
axially gauged coset
$SU(2)/U(1)_a$ or $SL(2)/U(1)_a$ and a free scalar field
at a compactification radius $r=0$.

As was already mentioned, the $R=1$ ($\a=0$) point,
corresponds to the original WZW model. Around this point, and for an
infinitesimal deformation parameter $\delta\a$, the deformed action
is given
by
\be
S_{R=1+\delta R}=S_{R=1}+{\delta R^2 \over 4\pi k}\int J\Jb\; ,
\ee
where $J$ and $\Jb$ are given in (\ref{JJ}).
This extends along the full line.
The $U(1)_{L}$ affine symmetry is generated by
$\th\rightarrow \th -\e$, $\tht\rightarrow \tht+\e /R^{2}$ with a
conserved
current
\be
J(R)=k\; {-(1-\S )\d\th +(1+\S )\d\tht\over 1+\S +R^{2}(1-\S)}\; ,
\ee
whereas the $U(1)_{R}$ affine symmetry is generated by
$\th\rightarrow \th +\epsilon$, $\tht\rightarrow\tht+\epsilon /R^{2}$
with
a conserved current
\be
\Jb(R)=k\; {(1-\S )\db\th +(1+\S )\db\tht\over 1+\S +R^{2}(1-\S)}\; .
\ee
Thus
\be
S_{R+\delta R}=S_{R}+{\delta R^{2}\over 4\pi k}\int J(R)\Jb (R)\;,
\label{pp}
\ee
and the variation of the dilaton provides the proper measure for the
$\sigma$-model above.

We now arrive to the important point of this section. The Weyl
transformation $J\rightarrow -J$ is given by a group rotation at the
WZW
point, and thus is a symmetry of the WZW model. Therefore, the
deformation
of the WZW model by $\delta\a J\Jb$ is equivalent infinitesimally to
the
deformation by $-\delta\a J\Jb$. The points $\delta\a$ and
$-\delta\a$
along the $\a$-modulus are thus the same CFT. In string theory, we
say
that they are related by a residual $Z_2$ symmetry of the broken
gauge
symmetry of the extended symmetry point.

The residual discrete symmetry is the target
space duality. This symmetry can be integrated to finite $\a$, giving
rise to a $Z_2\in O(2,2,Z)$ duality matrix $g_D$
\ba
g_D&=&
\left(\begin{array}{cccc} 0&I\\ I&0\end{array}\right)
\left(\begin{array}{cccc} I&-\e\\0&I\end{array}\right)
\left(\begin{array}{cccc} e_2&e_1\\ e_1&e_2\end{array}\right)
\left(\begin{array}{cccc} I&\e\\0&I\end{array}\right)
\left(\begin{array}{cccc} 0&I\\ I&0\end{array}\right)
\nonumber\\
&=&\left(\begin{array}{cccc}
{\left(\begin{array}{cccc} 0&1\\ 0&1\end{array}\right)} &
{\left(\begin{array}{cccc} 1&0\\ 0&0\end{array}\right)} \\
{\left(\begin{array}{cccc} 1&-1\\ -1&1\end{array}\right)} &
{\left(\begin{array}{cccc} 0&0\\ 1&1\end{array}\right)}
\end{array}\right),
\label{gD}
\ea
where
\be
e_1=\left(\begin{array}{cccc} 1&0\\ 0&0\end{array}\right), \qquad
e_2=\left(\begin{array}{cccc} 0&0\\ 0&1\end{array}\right),
\ee
$I$ is the 2-dimensional identity matrix, and $\e$ is given in
(\ref{e}).
The element $g_D$ acts on $E_R$ by (\ref{gE}) and gives
\be
g_D(E_R)=E_{1/R}.
\ee
Therefore, duality takes the modulus  $R$ to its inverse $1/R$.

As a consistency check, we note that $R\rightarrow 1/R$ is equivalent
to
$\tan\a\rightarrow\frac{-\tan\a}{1+k\tan\a}$, and therefore, a small
$\a$
is transformed by duality to $-\a$ as it should be.

In particular, we learn that the $R=0$ and $R=\infty$ points are the
same
CFT. Thus the vector/axial duality of
$SU(2)/U(1)$ and $SL(2)/U(1)$ is exact, and corresponds in string
theory to
a residual discrete symmetry of the broken gauge symmetry.

We should add some more comments here concerning equivalent versions
of the WZW deformations.
The first remark is that the background (\ref{ER},\ref{pR}) can be
obtained as
a gauged WZW model, $SU(2)\times U(1)/U(1)$ where the $U(1)$ has
radius
$e$ and the gauged $U(1)$ is the sum of the $U(1)$ of the free boson
and the
Cartan of $SU(2)$. Gauging the axial current and integrating out the
gauge
fields we obtain half of the line in (\ref{ER},\ref{pR}) with
$R^2=1+k/e^2$.
Gauging the vector current we obtain the other half of the line
with $R^2=(1+k/e^2)^{-1}$.
It is obvious that at $e\rightarrow\infty$ we are left with the WZW
model.
When $e\rightarrow 0$, before gauging, we know that the CFT is that
of a
non-compact boson times the WZW model.
The only compact $U(1)$ subgroup then is one that lies solely in
$SU(2)$
and thus we obtain the direct product of a non-compact boson times
the
$SU(2)/U(1)$ coset model.

The discussion in the Appendix shows that the background
(\ref{ER},\ref{pR})
is a conformally exact $\s$-model to all orders in a certain scheme.
In view of the relation to the coset above, this implies that there
is a scheme
in which the semiclassical result (obtained by integrating out the
gauge
fields) is exact to all orders.

Another dual version of our model can be obtained by performing a
factorized
duality transformation (\ref{factd}), followed by an appropriate
$GL(2,Z)$
transformation.
The action obtained thus is the sum of the parafermionic action and
the action
of a free scalar field with radius $\sqrt{k}R$, up to a $Z_{k}$
orbifoldization
which couples the two.
The orbifoldizing symmetry acts as a $Z_{k}$ transformation in the
parafermionic
theory and a simultaneous translation of the free scalar by $2\pi
/k$.
In this form of the action,
the factorization of the theory (at the boundary)
to that of a non-compact boson and the $SU(2)/U(1)$ parafermion
theory is
manifest:
When $R\to \infty$, the $Z_{k}$ translation acts trivially on the
non-compact
scalar, so the orbifolding symmetry acts uniquely in the parafermion
theory.
However, the parafermion theory is invariant under such an
orbifolding
\cite{GQ}.
This also supports the statement that the scalar with a vanishing
radius at the
boundary is a bona-fide non-compact scalar.
All of the above will be explicitly verified in the next section
by an analysis of the exact torus partition function.

\section{The partition function along the $R$-line}
\setcounter{equation}{0}

In this section we present the partition function, and discuss the
geometrical interpretation of the parameter $R$.

It is tempting to relate $R$ to the compactification radius of the
deformed
Cartan torus. This gets support by looking at the torus partition
function for such a deformation. The partition function is known in
the
$SU(2)_k$ deformed case, and is given by \cite{Y}
\be
Z(R)=\sum_{l,\bar{l}=0}^{k}\sum_{m=-k+1}^{k}\sum_{r=0}^{k-1}
N_{l,\bar{l}} c_m^l(q) \bar{c}_{m-2r}^{\bar{l}}(\bar{q})
\sum_{M,N\in Z} q^{\Delta_{M,N}}\bar{q}^{\bar{\Delta}_{M,N}}\;,
\label{ZR}
\ee
with
\be
\Delta_{M,N}=\frac{1}{4k}\left(\frac{kM+m-r}{R}+R(kN+r)\right)^2\;,
\ee
\be
\bar{\Delta}_{M,N}=\frac{1}{4k}\left(\frac{kM+m-r}{R}-R(kN+r)\right)^2
\;.
\ee
In (\ref{ZR}) $N_{l,\bar{l}}=\delta_{l,\bar{l}}$ (corresponding to
the
diagonal modular invariant WZW model), $c_m^l(q)$ are the standard
string
functions \cite{KP}, and $q=e^{2\pi i \tau}$, where
$\tau=\tau_1+i\tau_2$ is the complex torus
moduli parameter.

This partition function, indeed, can be obtained by performing a
$Z_{k}$
orbifold
on the direct tensor product of the $SU(2)/U(1)$ parafermion theory
and a free scalar field compactified on a circle of radius
$\sqrt{k}R$
\cite{Y}.
The $Z_{k}$ symmetry we orbifoldize with, is a combination
of the $Z_{k}$ parafermionic symmetry and a translation of the free
scalar
by $2\pi /k$.
This is explicitly demonstrated by writing (\ref{ZR}) as
\be
Z(R)={1\over k}\sum_{r,s\in Z_{k}}\zeta_{k}(r,s) Z(r,s,R)
\label{orb}
\ee
where $\zeta_{k}(r,s)$ is the parafermion partition function
twisted by the elements $r,s\in Z_{k}$ around the two non-trivial
cycles
and $Z(r,s,R)$ is the respective twisted partition function of the
free scalar
with radius $\sqrt{k}R$.
By doing the sum on $s$ in (\ref{orb}) we get the expression given in
(\ref{ZR}).

The partition function (\ref{ZR}) is invariant under the duality
transformation $R\rightarrow 1/R$. The fixed point $R=1$ corresponds
to the
WZW model.
Next we will check the two boundary points: $R=\infty$ and
$R=0$.

At $R\rightarrow \infty$ the only contribution to $Z$ comes when
$r=0$
and $N=0$ in the sums,
namely
\be
Z(R\rightarrow\infty)=\left(|\eta(q)|^2
\sum_{l,\bar{l}=0}^{k}\sum_{m=-k+1}^{k}
N_{l,\bar{l}} c_m^l(q) \bar{c}_{m}^{\bar{l}}(\bar{q})\right)
\left(\sqrt{k}R\tau_2^{1/2}|\eta(q)|^{-2}\right),
\label{Zinf}
\ee
where $\eta$ is the Dedekind eta-function.
The first parentheses  in (\ref{Zinf})
corresponds to the vectorially gauged coset \cite{K2}
\be
Z^v_{SU(2)/U(1)}=
|\eta(q)|^2
\sum_{l,\bar{l}=0}^{k}\sum_{m=-k+1}^{k}
N_{l,\bar{l}} c_m^l(q) \bar{c}_{m}^{\bar{l}}(\bar{q}),
\ee
which is the correct parafermionic partition function \cite{GQ}.
The second parentheses in (\ref{Zinf}) corresponds to a free
scalar field at the
decompactification limit or zero radius limit.

At $R\rightarrow 0$ the only three contributions to $Z$ come when
the following
conditions are obeyed in the sums in (\ref{ZR}): (1) $m-r=0$ and
$M=0$.  (2) $m-r=k$ and $M=-1$. (3) $m-r=-k$ and $M=1$.
Therefore, one finds
\ba
Z(R\rightarrow 0)&=&\left[|\eta(q)|^2
\sum_{l,\bar{l}=0}^{k}
N_{l,\bar{l}}\left(\sum_{m=-k}^{-1} c_m^l(q)
\bar{c}_{-m-2k}^{\bar{l}}
(\bar{q})
+\sum_{m=0}^{k-1}c_{m}^{l}(q)
\bar{c}_{-m}^{\bar{l}}(\bar{q})
+c^{l}_{k}(q)\bar c^{\bar l}_{k}(\bar{q})\right)\right]\nonumber\\
&{}&\times\left(\sqrt{k}R^{-1}\tau_2^{1/2}|\eta(q)|^{-2}\right).
\label{Z0}
\ea
The first parentheses  in (\ref{Z0})
corresponds to the axially gauged coset \cite{K2}
\be
Z^a_{SU(2)/U(1)}=|\eta(q)|^2
\sum_{l,\bar{l}=0}^{k}
N_{l,\bar{l}}\left(\sum_{m=-k}^{-1} c_m^l(q)
\bar{c}_{-m-2k}^{\bar{l}}
(\bar{q})+\sum_{m=0}^{k-1}c_{m}^{l}(q)
\bar{c}_{-m}^{\bar{l}}(\bar{q})
+c^{l}_{k}(q)\bar c^{\bar l}_{k}(\bar{q})\right).
\ee
Using the symmetry of the string functions $c_m^l$ under the affine
Weyl
group one obtains that \cite{K2}
\be
Z^a=Z^v.
\ee
The second parentheses in (\ref{Z0}) corresponds
to a free scalar field at the
decompactification limit or zero radius limit.

We  conclude that the partition function (\ref{ZR}) corresponds to
the
deformation line of the $SU(2)$ WZW background described in this
section.
The parameter $R$ in (\ref{ER}) is therefore related to the radius of
the
Cartan torus. Although we do not know the exact partition function of
$SL(2,R)$ for arbitrary Cartan radius, our previous argument implies
that
similar things happen also there.

\section{Geometry along the $R$-line and smooth topology change}
\setcounter{equation}{0}

In this section we describe the geometry along the $R$-line
and a smooth topology change in the extended moduli
space of $SU(2)$ or $SL(2)$.

We will first present the description of the
target-space geometry along the $R$-line of marginal deformations.
The $\sigma$-model metric in the case of deformed
$SU(2)$ (in the coordinates $\th$, $\tht$, $x$)
is given by
\be
G\sim k\left(
\matrix{{\sin^2 x\over  \cos^2 x +R^2 \sin^2 x}& 0&0\cr
0&{R^{2}\cos^2 x\over \cos^2 x+R^2 \sin^2 x }&0\cr
0&0&1\cr}\right).
\ee
The scalar curvature ${\hat R}$ is
\be
{\hat R}=-{2\over k}{2-5R^2 +2(R^{4}-1)\sin^2 x \over
 (1+(R^2-1)\sin^2 x)^2} \; .
\label{r}
\ee
The manifold is regular except at the end-points where
\be
{\hat R}(R=0)=-{4\over k\cos^2 x}\;\;,\;\;{\hat R}(R=\infty)=-{4\over
 k\sin^2 x}\; .
\ee
At $R=1$ we get the constant curvature of $S^{3}$, ${\hat R}=6/k$.

It should be noted that the geometric data (metric, curvature, etc.)
are
invariant under $R\rightarrow 1/R$ and $x\rightarrow \pi/2-x$.
Another interesting object is the volume of the manifold as a
function of $R$
that can be computed to be
\be
V(R)\sim {R\;{\log}R\over R^{2}-1}
\ee
satisfying $V(R)=V(1/R)$. The volume becomes singular
only at the boundaries of moduli space, $R=0,\infty$.

For $SL(2,R)$, the trigonometric functions in (\ref{r}) are replaced
by the
corresponding hyperbolic functions. Here the manifold has a curvature
singularity for $0\leq R <1$. Similar remarks apply to the Euclidean
version, the 3-d hyperboloid.

The $R$ marginal deformation generates a continuous  family of CFTs
that
interpolate between two manifolds with  different topology: the
``cigar"
shape ($R=\infty$) with the topology of the disk and the ``trumpet"
shape
($R=0$) with the topology of a cylinder. In the $SU(2)$ case, the
$S^3$
group manifold is deformed to  $D_2\times p$, i.e., the direct
product of a two disc and a point. These topology changes  occur only
at the boundary of moduli space; a smooth topology change will be
described
below.

In the $SU(2)$ case,  the complete moduli space of backgrounds also
includes target spaces that are a direct product of $SU(2)/U(1)$ like
backgrounds (namely, $ds^2=kdx^2+k'\tan^{\pm 2}x d\th^2$, where $k'$
is
an arbitrary constant) with a {\it
finite} radius circle $S^1$. These backgrounds are smooth
deformations of
the $SU(2)$ group manifold $S^3$. For example, we can rotate the
$SU(2)$
background matrix  $E$ in (\ref{Ssu2},\ref{Esu2}) with the  $O(2,2)$
element
\be
\tilde{g}_{\a}=
\left(\begin{array}{cccc} I&k\e\\0&I\end{array}\right)
\left(\begin{array}{cccc} C(\a)&S(\a)\\S(\a)&C(\a)\end{array}\right)
\left(\begin{array}{cccc} I&-k\e\\0&I\end{array}\right),
\ee
where $\e, C(\a), S(\a)$ are given in (\ref{e}).
One finds
\be
\tilde{g}_{\a}(E)=\frac{k}{\Delta}
\left(\begin{array}{cccc} 1-\S & B \\-B & 1+\S \end{array}\right),
\label{Eline}
\ee
where
\ba
\S=\cos 2x\; , &{}&\qquad
\Delta=\cos^2 \a (1+\S )+(\cos\a+k\sin\a)^2(1-\S )\; ,\nonumber\\
B&=&\frac{1}{k}\sin(2\a)-(\cos(2\a)+\frac{k}{2}\sin(2\a))(1-\S)+\Delta
 \; .
\ea
The dilaton transforms by eq. (\ref{pR}).

The metric along the $\a$ line of deformations is then given by the
line element
\be
ds^2(\a)=\frac{k}{\Delta(\a)}[(1-\S)d\th^2+(1+\S)d\tht^2]+kdx^2.
\label{line}
\ee
At the point $\a=0$ the background (\ref{Eline},\ref{line})
is the $SU(2)_k$ group manifold $S^3$ (with an antisymmetric
background).
Along the line $0<\a <\pi/2$ the background (\ref{Eline}) includes
the
metric (\ref{line}) with the topology
of $S^3$, as well as an antisymmetric background and a dilaton field.
At the point $\a=\pi/2$ the background metric is
\be
ds^2(\a =\pi/2)=\frac{1}{k}[d\th^2+\frac{1+\S}{1-\S}d\tht^2 ]+kdx^2.
\ee
At this point the manifold has a topology of $D_2\times S^1_{1/k}$,
where $D_2$
is a two disc and $S^1_{1/k}$ is a circle with radius $r^2=1/k$.
One may continue to deform this theory by, for example, changing the
compactification radius $r$ of the free boson $\th$.

Therefore, we find that the {\it quantum theories} based on
$\sigma$-models with topologically distinct target spaces in the
extended
moduli space of the WZW model are  smoothly connected, even though
classically a physical singularity is encountered.

It is remarkable that (for integer $k$)  the neighborhood of the
point
$\a=\pi/2$ is mapped to the neighborhood of the point $\a=0$ by an
element
of $O(2,2,Z)$, namely, a target space generalized duality
\cite{RV,GR,Kumar}.  Therefore, a region in the moduli space where a
topology change occurs is mapped to a region where there is no
topology
change at all. This is very similar to the observation made in the
Calabi-Yau case \cite{Getal,Witten}, where mirror symmetry  plays the
same
role as $O(2,2,Z)$ plays here.

\section{$J\Jb$ deformations in curved backgrounds}
\setcounter{equation}{0}

In this section we describe $J\Jb$ deformations  of general WZW
models. The
relation of these deformations to $O(d,d)$ transformations was
described
infinitesimally in \cite{K2} and in finite form in \cite{HN}. We will
start
with a particular parametrization of a WZW model for a group $G$. The
group
$G$ can be semisimple, although we will explicitly  indicate one of
the
levels $k$, while the others are hidden in the action. The relevant
level
is the one corresponding to the simple component of the group whose
Cartan
we are deforming. The following parametrization can be easily
obtained
using the Polyakov-Wiegmann formula (see for example \cite{K,K2})
\be
S[x^a,\tha,\thb]=S_1+S_a+S[x],
\label{wzwbeg}
\ee
\be
S_1=\frac{k}{2\pi}\int d^2 z \left(\d\tha \db\tha + \d\thb \db\thb
+2\S (x) \d\thb \db\tha+\G_a^1(x)\d x^a\db\tha+\G_a^2(x)\d\thb\db
x^a\right),
\ee
\be
S_a=\frac{k}{2\pi}\int d^2 z(\d\thb\db\tha - \d\tha\db\thb),
\ee
\be
S[x]=\frac{k}{2\pi}\int d^2 z\;\G_{ab}(x)\d x^a \db x^b -
\frac{1}{8\pi}\int d^2 z\;\p_0 R^{(2)},
\label{wzw}
\ee
where $a=1,...,D$, $D={\rm dim}G-2$, and $\p_0$ is a constant
dilaton. The
backgrounds $\S(x), \G^1(x), \G^2(x)$ and $\G(x)$ are
independent of the coordinates $\tha,\thb$.
With the coordinates $\th$ and $\tht$ defined in (\ref{th},
\ref{th1}), the
action becomes
\be
S[x^a,\th,\tht]=\frac{1}{2\pi}\int d^2 z
(\d\th, \d\tht, \d x^a)\left( \begin{array}{cccc}
                                          E & F_b^2\\
                                          F_a^1 & F_{ab}
                                         \end{array}\right)
\left(\begin{array}{clcr} \db \th \\ \db \tht \\ \db x^b\end{array}
\right)
-\frac{1}{8\pi}\int d^2 z\;\p_0 R^{(2)},
\label{Swzw}
\ee
where the $2\times 2$ background matrix $E$, the
$D\times 2$ matrix $F_b^2$, the $2\times D$ matrix
$F_a^1$, and the $D\times D$ matrix $F_{ab}$ are given by
\be
\left(\begin{array}{clcr}
                             E & F_b^2\\
                             F_a^1 & F_{ab}
                             \end{array}\right)=
\frac{k}{2}
\left(\begin{array}{clcr}
{\left(\begin{array}{clcr} 1-\S & 1+\S\\
      -(1+\S) & 1+\S \end{array}\right)} & {\left(\begin{array}{clcr}
                                           \G_b^2 \\ \G_b^2
                                            \end{array}\right)}\\
{\left(-\G_a^1 \;\;\;\;\;\;\; \G_a^1 \right)} & 2\G_{ab}
\end{array}\right).
\label{EFFF}
\ee

The action (\ref{Swzw}) is manifestly invariant under the
$U(1)_L\times
U(1)_R$ affine symmetry generated by the currents
\ba
J=\frac{k}{2}\left(-(1-\S)\d\th +(1+\S)\d\tht+\G_a^1\d x^a\right),
\nonumber\\
\Jb=\frac{k}{2}\left((1-\S)\db\th +(1+\S)\db\tht+\G_a^2\db
x^a\right).
\label{JJb}
\ea
In addition, there are extra ${\rm dim}G-1$ chiral currents and
${\rm dim}G-1$ anti-chiral currents, completing the affine $G_L\times
G_R$
symmetry of the WZW model.~\footnote{
Actually, one can bring the background in (\ref{wzwbeg})-(\ref{wzw})
into the form appearing in ref.
\cite{GR} eq. (2.1) with $i=1,...,{\rm rank}G$. In this form
the ${\rm rank}G$ chiral currents and ${\rm rank}G$ anti-chiral
currents
corresponding to the left-handed and right-handed Cartan tori are
manifest.}

We now deform the action (\ref{Swzw}) with $J\Jb$, where $J$ and
$\Jb$ are
given in (\ref{JJb}).  The $U(1)$ chiral and anti-chiral currents at
the
deformed theory will be found.

The deformation with $J\Jb$ is equivalent to
$g\in O(2,2)$ rotations acting on the background matrices $E, F^1,
F^2, F$.
The action of $g$ in (\ref{gJ})
on $E$ is given in (\ref{gE}). Here we need also the
action of $g$ on the background matrices $F^1,F^2, F$ \cite{GR}:
\ba
g(F^1)=F^1(cE+d)^{-1}, \qquad g(F^2)=(a-E'c)F^2,\nonumber\\
g(F)=F-F^1(cE+d)^{-1}cF^2,
\label{gF}
\ea
where $a,c,d$ are defined in (\ref{gJ}), and $E'$ is given in
(\ref{gE}).

The one parameter family of the
$J\Jb$ deformation is given by
acting on the background (\ref{EFFF}) with $g_{\a}$ given in
(\ref{ga}).
Using eqs. (\ref{gE}, \ref{gF}) one finds
\be
\frac{1}{k}g_{\a}(E)\equiv E_{R(\a)},
\ee
\be
\frac{1}{k}g_{\a}(F^1)\equiv F^1_{R(\a)}
=\frac{(1+\S)^{-1}\G^1}{1+R^2\frac{1-\S}{1+\S}}
\left(-1,R^2\right),
\label{F1R}
\ee
\be
\frac{1}{k}g_{\a}(F^2)\equiv F^2_{R(\a)}
=\frac{(1+\S)^{-1}\G^2}{1+R^2\frac{1-\S}{1+\S}}
\left(\begin{array}{clcr} 1 \\ R^2 \end{array}\right),
\label{F2R}
\ee
\be
\frac{1}{k}g_{\a}(F)\equiv F_{R(\a)}
=\G+\frac{(R^2-1)(1+\S)^{-1}\G^1\G^2}{2(1+R^2\frac{1-\S}{1+\S})},
\label{FR}
\ee
where  $E_R$ and $R(\a)$ are given in (\ref{ER}, \ref{R}).

The constant dilaton $\p_0$ in the WZW background (\ref{wzw})
transforms
under $g_{\a}$ by eq. (\ref{pR}). For more details see ref.
\cite{GR}.

{}From eqs. (\ref{ER}, \ref{F1R}, \ref{F2R}, \ref{FR}), we see that
up to an overall $k$ factor, the background is parametrized by one
radius $R^2 =(1+k\tan \a)^2$.

As in the $SU(2)$ and $SL(2)$ cases,
the whole Cartan subalgebra survives along the deformation.
In the parametrization we are using we have only one pair of these
explicit
\be
J(R)=k\;{{-(1-\S)\d\th+(1+\S)\d\tht+\Gamma_{a}^{1}\d x^{a}}\over
 1+\S+R^{2}(1-\S)}\; ,
\ee
\be
{\bar J}(R)=k\;{{(1-\S)\db\th+(1+\S)\db\tht+\Gamma_{a}^{2}\db
x^{a}}\over
 1+\S+R^{2}(1-\S)}\; ,
\ee
and we can verify that that eq. (\ref{pp}) is still valid.
In fact it does not matter with which current in the Cartan of a
simple
component we are deforming. Different choices are related by target
space
reparametrizations.

It is instructive here to discuss the counting of parameters of
$O(d,d)$
transformations related to that of marginal deformations. A general
WZW
model for a semi-simple group $G$ has $2r$ Killing symmetries
associated
with the currents of the Cartan subalgebra of $G_L\times G_R$, where
$r$ is
the rank of $G$. Thus the relevant group is $O(2r,2r)$. $GL(2r)$
transformations and antisymmetric tensor shifts preserve the presence
of
the current algebra, although they might change its spectrum. Out of
this
($6r^{2}-r$)-dimensional subgroup no transformation preserves the
action.
There is a $r(2r-1)$-dimensional manifold of left-over
transformations
which break the $G$-current algebra; these are of the type $J\Jb$.
Therefore, the  $J\Jb$ deformations correspond to  a $r(2r-1)$
parameter
family of the full $O(2r,2r)$ group. Out of these, the
$r(r-1)$-dimensional subgroup $O(r)\times O(r)$  leaves the action
invariant. The rest $r^2$ transformations correspond precisely to all
possible $J\Jb$ marginal perturbations with currents in the Cartan
subalgebra.

The deformation described here is true for any background with chiral
and
anti-chiral currents, and it is only for the purpose of discussing
gauge
symmetries that we assume that some point on this line (in our
notation
$R=1$ ) is a WZW model.

In the next section we discuss target space duality on the $R$-line
of
deformations.

\section{Duality on the $J\Jb$ line is a residual broken gauge
symmetry}
\setcounter{equation}{0}

Remarkably,  the action of $g_D\in O(2,2,Z)$
given in eq. (\ref{gD}) on the backgrounds $E,F^1,F^2,F$ in
eq. (\ref{EFFF}), gives a rather simple duality transformation on the
$R$-modulus. By
straightforward calculations using
the transformations given in (\ref{gE}, \ref{gF}) one
finds
\be
g_D\left(
\left(\begin{array}{cccc}
                             E_R & F_R^2\\
                             F_R^1 & F_R
                             \end{array}\right)\right)=
\left(\begin{array}{cccc}
                             E_{1/R} & F_{1/R}^2\\
                             F_{1/R}^1 & F_{1/R}
                             \end{array}\right).
\ee
Therefore, duality takes the modulus $R$ to its inverse $1/R$.

The results described in section 2 for the $SU(2)$ and $SL(2)$ cases
are
extended to the general case. The fixed point $R=1$ corresponds to
the
extended symmetry point $G_L\times G_R$, namely, the original WZW
model.
Infinitesimally around the extended symmetry point, duality
corresponds to
the transformation $\a\rightarrow -\a$. This transformation can be
achieved
by a Weyl transformation  in $G_L$  (or $G_R$) that reflects
$J\rightarrow
-J$ (or $\Jb\rightarrow -\Jb$). Duality is related to a Weyl
reflection and
is, therefore, a residual discrete symmetry of the broken gauge
algebra of
the associated string theory. This provides a map between the two
half
lines, and in particular identify the two boundaries at $R=0$ and
$R=\infty$. In the following we show that these boundaries
correspond,
respectively, to the direct product of the cosets $G/U(1)_a$ and
$G/U(1)_v$
with a free scalar field.

At $R=0$ ($\a=tan^{-1}(-\frac{1}{k}$)) the background matrix is
\ba
E_{R=0}=
\left(\begin{array}{cccc} \frac{1-\S}{1+\S}&1\\-1&0\end{array}\right)
&=&
\left(\begin{array}{cccc} \frac{1-\S}{1+\S}&0\\0&0\end{array}\right)
\;\;\; {\rm mod} \;\; \Theta - {\rm shift},
\nonumber\\
F_{R=0}^1=(1+\S)^{-1}\G^1(-1,0),&{}& \qquad
F_{R=0}^2=(1+\S)^{-1}\G^2\left(\begin{array}{clcr}
1\\0\end{array}\right),
\nonumber \\
F_{R=0}&=&\G-\frac{1}{2}(1+\S)^{-1}\G^1\G^2.
\label{EFR0}
\ea
This background corresponds \cite{RV,GR}
to the direct product of the vectorially gauged coset
$G/U(1)_v$ and a free boson at a compactification
radius $r=0$ (that is equivalent to a non-compact free boson via the
$r\rightarrow 1/r$ duality).

At $R=\infty$ ($\a=\pi/2$) the background matrix is
\ba
E_{R=\infty}=
\left(\begin{array}{clcr} 0&0\\ 0&\frac{1+\S}{1-\S}\end{array}\right)
,
&{}& F_{R=\infty}=\G+\frac{1}{2}(1-\S)^{-1}\G^1\G^2 ,
\nonumber\\
F_{R=\infty}^1=(1-\S)^{-1}\G^1(0,1), \qquad
&{}& F_{R=\infty}^2=(1-\S)^{-1}\G^2
\left(\begin{array}{clcr} 0\\1\end{array}\right) .
\label{EFRinf}
\ea
This background corresponds \cite{RV,GR}
to the direct product of the axially gauged coset
$G/U(1)_a$ and a free boson at a compactification
radius $r=0$.

Therefore, like the $SU(2)$ and $SL(2,R)$ cases,  because the
end-points are equivalent theories we obtain that
axial-vector duality is exact in general.

\section{Comments and open problems}
\setcounter{equation}{0}

In this work we have studied some aspects of the global structure of
the
moduli space of curved string backgrounds with $d$ toroidal
isometries.
The moduli space of such  CFTs (at the $\sigma$ model level )
contains the
moduli space that is generated by $O(d,d,R)$ transformations. The
latter
contains all marginal deformations of the $J\Jb$ kind, with $J$,
$\bar J$ in
the Cartan. The symmetries
acting on the $O(d,d,R)$ space are generated by the $O(d,d,Z)$ group.
What
we have shown in this work is that some elements of this group,
corresponding to axial-vector duality of abelian cosets of WZW
models,  can
be viewed as residual gauge symmetries since around a WZW point they
are
related by a continuous group transformation.  Do all the $O(d,d,Z)$
transformations correspond to  residual gauge symmetries? This
question was
addressed in the flat case \cite{GMR2} where it was shown that any
$O(d,d,Z)$ transformation corresponds to a product of discrete
residual
gauge symmetries around  points with extended symmetry.

At the $\sigma$-model level, the deformations given by $O(d,d)$
transformations are not the full story; one can generate more
(equivalent)
backgrounds by deforming, for example,  with $J\Jb$ combinations
different
from the ones considered here (namely, of the type
$(\sum_{a=1}^{a=D}
\a_{a}J^a)(\sum_{b=1}^{b=D}\beta_b \Jb^b)$,  where $D={\rm dim} G$).
Although these are related by transformations in the $G_{L}\times
G_{R}$
symmetry group, they give $\sigma$-models which are not related to
the
previous one by a coordinate transformation (unless the
transformation is
in the diagonal $G$). A similar effect appears when gauging $U(1)$
sub-groups, that are embedded differently in the group, giving rise
to a
family of  different actions for the same coset.

The deformation line of the $SL(2)$ model described in section 2
interpolate between  the {\it  euclidean}  abelian cosets. Therefore,
we
proved duality for the euclidean $SL(2)/U(1)$ coset.  A proof for the
lorentzian case proceeds along the same lines by deforming
with the $J\Jb$ corresponding to the non-compact $U(1)$ in $SL(2)$.

A final remark concerns ``topology change". The marginal deformations
discussed in this paper provide some concrete examples of continuous
change
of topology by relating the axial to vector cosets, and to the
original
$WZW$ points. We found that a smooth topology change occurs in the
extended
moduli space of $SU(2)_k$ model. This can be extended to the moduli
space
of $SL(2)_k$, and other curved string backgrounds with one time-like
coordinate; it therefore provides an evidence to a {\it classical
topology
change in cosmological string backgrounds}  (and other curved
space-time).
Moreover, in these cases the different topologies might give rise to
equivalent
theories, a phenomenon that is not unheard of in string theory.

\vskip 1cm
\noindent
{\bf Acknowledgments} \\

\noindent
We thank S. Elitzur, C. Kounnas, A. Polychronakos, E. Rabinovici, M.
Ro\v cek
and A. Shapere
for discussions. We also thank C. Kounnas and M. Ro\v cek for useful
comments
on the
paper. AG would like to thank the Institute for theoretical Physics
at
Santa Barbara where this work was completed. EK would like to thank
the Tel
Aviv University and especially the  Hebrew University in Jerusalem
for
their warm hospitality while this work was done. This work was
supported in
part by NSF grant No. PHY89-04035.
We would also like to thank the referee for prompting us to make our
arguments clearer.

\vskip 1cm
\noindent
{\bf Note Added}\\
\vskip .3cm

In \cite{Moo}, Moore gave a precise prescription of Conformal
Perturbation
Theory which preserves
the duality symmetry $R\rightarrow 1/R$ in the theory of compact
scalar fields.
It is straightforward to show that precisely the same prescription
does the job
here, both in the compact and non-compact theories, since an
arbitrary
correlation function factorizes into that of a free boson and a
parafermionic
one, which does not feel the perturbation.
The interpretation of this result is that the duality symmetry is
non-anomalous
in perturbation theory.
We thank the referee for bringing this to our attention.

In \cite{Tsey}, Tseytlin showed that there is a scheme in $\s$-model
perturbation theory where the semiclassical background for the
(compact or non-compact) parafermion theory is exact to all orders.
Our arguments show that such a scheme exists for the whole line of
theories
and not only for the boundary or WZW points.
The above make the following conjecture highly plausible: For all
coset
models (compact or non-compact) there is a scheme where the
semiclassical
background (obtained from the gauged WZW model) is exact.

\newpage
\setcounter{section}{0}
\setcounter{equation}{0}

\renewcommand{\thesection}{Appendix \Alph{section}.}
\renewcommand{\theequation}{A.\arabic{equation}}

\section{}

In this Appendix we will give the detailed argument concerning
the exactness of the $\sigma$-model picture.
Conformal perturbation theory (CPT) indicates that there is a line of
theories
obtained by perturbing around the $SU(2)$ or $SL(2)$ WZW model  by
$\int J^{3}{\bar J}^{3}$.
The theories along the line have a $U(1)_{L}\times U(1)_{R}$ chiral
symmetry.
Thus the $\sigma$-model action of these theories must satisfy at
least
the following three properties

1) It should have $U(1)_{L}\times U(1)_{R}$ chiral symmetry along the
line.

2) It should have the group property: $\delta S \sim \int  J^{3}{\bar
J}^{3}$
at {\em any} point of the line (this is a property obvious in CPT).

3) At a specific point it should reduce to the known action of the
   $SU(2)$ or $SL(2)$ WZW model.

The most general 3-d action which satisfies property (1) is \cite{RV}
\be
S[x,\psi_{1},\psi_{2}]=S_1+S_2+S_3,
\label{w1}
\ee
\be
S_1=\frac{1}{2\pi}\int d^2 z \left(\d\tha \db\tha + \d\thb \db\thb
+2\S (x,\lambda) \d\tha \db\thb+\G_1(x,\lambda)\d
x\db\thb+\G_2(x,\lambda)\d\tha\db
x\right),
\ee
\be
S_2=\frac{1}{2\pi}\int d^2 zB(\lambda)(\d\tha\db\thb - \d\thb\db\tha)
+\frac{1}{2\pi}\int d^2 z\;\G(x,\lambda)\d x \db x,
\ee
\be
S_3=-\frac{1}{8\pi}\int d^2 z\;\p (x, \lambda) R^{(2)},
\label{w2}
\ee
where
\be
\theta_{i}=\sum_{j=1}^{2}\alpha_{ij}(\lambda)\psi_{j},
\ee
and we have explicitly indicated the dependence on the continuous
parameter
$\lambda$.
Here the $\psi_{i}$ are $\lambda$-independent angular coordinates.

The chiral currents can be calculated to be
\be
J(\lambda)=\d\thb+\S (x,\lambda)\d\tha+{1\over 2}\G_1(x,\lambda)\d x
\ee
\be
{\bar J}(\lambda)=\db\tha+\S (x,\lambda)\db\thb+{1\over
2}\G_2(x,\lambda)\db x
{}.
\ee

We now impose property (2) to $S_{1}+S_{2}$ \footnote{
The dilaton should be considered as part of the measure of the path
integral
since its contribution is visible at the one loop level.
It should of course be determined and we will do so in two
independent
ways, either from the requirement that the measure is correct
along the line, as it was explained in the main text, or just from
the
requirement of conformal invariance that we will use here.}:
\be
{\d\over \d \lambda}S_{1+2}(\lambda )={g(\lambda)\over 2\pi}\int
J(\lambda){\bar J}
(\lambda),\label{de}
\ee
where $g(\lambda)$ reflects the freedom of independent normalization
of the
currents along the line.
However, it can always be set to any fixed number by a
reparametrization in
$\lambda$.
Thus, from now on, we will assume without loss of generality that
$g=2$.
Then (\ref{de}) gives a system of first order non-linear differential
equations which can be integrated explicitly.
Imposing also the boundary conditions:
\be
\a_{ij}(0)=\left(\matrix{\sqrt{k}&0\cr 0&\sqrt{k}\cr}\right),
\ee
\be B(0)=k\;\;,\;\;\S(x,0)=\S (x),
\ee
\be
\G_{1}(x,0)=\G_{2}(x,0)=0\;\;,\;\;\G(x,0)=k,
\ee
we obtain
\be
\a_{ij}(\lambda)=\sqrt{k}\left(\matrix{\cosh(\lambda)&\sinh(\lambda)
\cr
\sinh(\lambda)&\cosh(\lambda)\cr}\right),
\ee
\be
B=k\;\;,\;\;\Sigma(x,\lambda)={1-{1-\S(x)\over
1+\S(x)}e^{-2\lambda}\over
1+{1-\S(x)\over 1+\S(x)}e^{-2\lambda}},
\label{sis}
\ee
\be
\G_{1}(x,\lambda)=\G_{2}(x,\lambda)=0\;\;,\;\;\G(x,\lambda)=k.
\ee
This is precisely the background (\ref{ER}) with the identification
$R=e^{-\lambda}$.

So far we have not imposed conformal invariance.
At one-loop the $\beta$-function equations amount to (prime indicates
differentiation with respect to $x$)
\be
\p=-\log[\S'/\sqrt{1-\S^{2}}]+{\rm constant}
\ee
and
\be
\left({\S''\over \S'}\right)'+{2\S\S''+\S'^{2}\over
1-\S^{2}}+3{\S^{2}\S'^{2}\over (1-\S^{2})^{2}}=0\label{beta}
\ee
It is important to note that with $\S(x)=\cos(2x)$ ($SU(2)$) or
$\cosh(2x)$
($SL(2)$)
(\ref{sis}) satisfies the one-loop equation (\ref{beta}).
The only way (\ref{sis}) can change consistent with our requirements
(1-3)
is by a redefinition of $R$, which implies that there is a scheme in
$\sigma$-model perturbation theory where the metric and the
antisymmetric tensor
receive no higher order corrections in $\a'$.
In such a case also the dilaton receives no higher order corrections
\cite{tse}\footnote{The dilaton $\beta$-function (central charge)
does get corrections.
This is what is happening also in the WZW model.
However, like that case, one can replace $k$ with $k+2$ in front of
the action.
Then the central charge is given by the classical and 1-loop piece
only, without spoiling the vanishing of the other
$\beta$-functions.}.

Consequently, the dilaton is given by
\be
\p(x,R)=\log[1+{1-R^2\over 1+R^2}\S(x)]+f(R).
\ee
The $R$-dependent constant can be fixed by the requirements that
$\sqrt{G(R)}e^{\phi(R)}$ (which represents the physical string
coupling) is invariant along the line.
This gives the formula for the dilaton presented in section 2.

The arguments above show that (\ref{ER},\ref{pR}) describe the
correct
(to all orders in $\a'$)
$\sigma$-model background associated with the deformation of the WZW
model.
One further comment is in order to ensure that our conclusions hold
non-perturbatively in $\a'$.
The correlators of a generic theory along the $R$-line (compact or
not)
are products of a parafermionic correlator coupled to a block of a
boson
at radius $R/\sqrt{k}$.
For the boson blocks we know their explicit structure.
What remains to analyse is the non-perturbative structure of the
parafermionic blocks (which is the same as that of the respective
WZW).
In the compact case there are non-perturbative contributions which
can be seen to come from the non-trivial affine null vectors.
A look at the exact torus partition function (3.1) suffices to note
that the non-pertrubative terms are precisely those coming from the
subtraction
of affine null vectors (with dimensions of ${\cal O}(k)$).
This persists for partition functions at any genus.
An independent look at the sphere four-point amplitudes confirms
again that the non-perturbative terms come from the part of the
overall
coefficient which enforces the affine cutoff \cite{zinf}(which we
know to be the consequence of affine null vectors).
Such non-perurbative corrections do not spoil the $R\rightarrow 1/R$
duality.
In the non-compact case no non-perturbative corrections are expected
since there are no non-trivial affine null vectors.
A computation of the partition function in this case at $R=1$
corroborates this
statement \cite{Ga}.


\newpage

\begin{thebibliography}{6666}
\bibitem{nsw} K. Narain, M. Sarmadi and E. Witten, Nucl. Phys. {\bf
B279}
(1987) 369.

\bibitem{moduli} K.A. Meissner and G. Veneziano, Phys. Lett.
                {\bf B267} (1991) 33;\\
      M. Gasperini, J. Maharana and G. Veneziano,
      Phys. Lett. {\bf B272} (1991) 277; \\
      A. Sen, Phys. Lett. {\bf B271}
      (1991) 295.

\bibitem{GR} A. Giveon and M. Ro\v{c}ek, Nucl. Phys. {\bf B380}
(1992) 128.

\bibitem{GRV} A. Giveon, E. Rabinovici and G. Veneziano, Nucl. Phys.
              {\bf B322} (1989) 167;\\
              A. Shapere and F. Wilzcek, Nucl. Phys. {\bf B320}
             (1989) 669; \\
             A. Giveon, N. Malkin and E. Rabinovici, Phys. Lett.
             {\bf B220} (1989) 551.

\bibitem{K}   E.B. Kiritsis, Mod. Phys. Lett. {\bf A6} (1991) 2871.

\bibitem{Getal} P.S. Aspinwall, B.R. Greene and D. Morrison,
Phys.Lett. {\bf B303} (1993) 249.

\bibitem{Witten} E. Witten, Nucl. Phys. {\bf B403} (1993) 159.

\bibitem{DHS} M. Dine, P. Huet and N. Seiberg, Nucl. Phys. {\bf B322}
            (1989) 301.

\bibitem{GMR2} A. Giveon, N. Malkin and E. Rabinovici, Phys. Lett.
              {\bf B238} (1990) 57.

\bibitem{GP} A. Giveon and M. Porrati, Phys. Lett. {\bf B246} (1990)
54;\\
             A. Giveon and M. Porrati, Nucl. Phys. {\bf B355} (1991)
422.

\bibitem{KZ} T.Kugo and B. Zwiebach, Prog. Theor. Phys. {\bf 87}
(1992)
801.

\bibitem{G} A. Giveon, Nucl. Phys. {\bf B391} (1993) 229.

\bibitem{IL} L. Ibanez and D. L\"ust,  Phys.Lett. {\bf B302} (1993)
38.

\bibitem{K2} E. Kiritsis, Nucl. Phys. {\bf B405} (1993) 109.

\bibitem{GDVV} A. Giveon, Mod. Phys. Lett. {\bf A6} (1991) 2843;\\
R. Dijkgraaf, E. Verlinde, and H. Verlinde, Nucl. Phys. {\bf B371}
(1992)
269.

\bibitem{Ga} K. Gawedski, Lectures given at the Cargese Summer
Institute ``
New Symmetry Principles in QFT", July 1991.

\bibitem{HS} S.F. Hassan and A. Sen, Nucl. Phys. {\bf B405} (1993)
143.

\bibitem{brc} K. Bardakci, M. Crescimanno and E. Rabinovici, Nucl.
Phys. {\bf
B344} (1990) 344.

\bibitem{GQ} D. Gepner and Z. Qiu, Nucl. Phys. {\bf B285} (1987) 423.

\bibitem{Y} S.K. Yang, Phys. Lett. {\bf B209} (1988) 242.

\bibitem{KP} V. Ka\v{c} and D. Peterson, Adv. Math. {\bf 53} (1984)
125.

\bibitem{RV} M. Ro\v{c}ek and E. Verlinde, Nucl. Phys. {\bf B373}
(1992) 630.

\bibitem{Kumar} A. Kumar, Phys. Lett. {\bf B293} (1992) 49.

\bibitem{HN} M. Henningson  and C. Nappi, Phys.Rev. {\bf D48} (1993)
861.

\bibitem{Moo} G. Moore, ``{\em Finite in All Directions}",
hepth/9305139.

\bibitem{Tsey} A. Tseytlin, ``{\em On Field Redefinitions and Exact
Solutions
in String Theory}", hepth/9308042.

\bibitem{tse} R. Metsaev and A. Tseytlin, Nucl. Phys. {\bf B293}
(1987) 385.

\bibitem{zinf} I. Bakas and E. Kiritsis, Nucl. Phys. {\bf B343}
(1990) 343.

\end{thebibliography}
\end{document}